\documentclass{article}

\usepackage{arxiv}

\usepackage[utf8]{inputenc} 
\usepackage[T1]{fontenc}    
\usepackage{hyperref}       
\usepackage{url}            
\usepackage{booktabs}       
\usepackage{amsfonts}       
\usepackage{nicefrac}       
\usepackage{microtype}      
\usepackage{lipsum}		
\usepackage{graphicx}
\usepackage{natbib}
\usepackage{doi}

\usepackage{epstopdf, epsfig}
\usepackage{placeins}
\usepackage{dcolumn}
\usepackage{bm}
\usepackage[dvipsnames]{xcolor}
\usepackage{comment}
\usepackage{hyperref}
\usepackage{amsmath}

\title{Universality of stretching separation}

\author{ \href{https://orcid.org/0000-0003-2505-3742}{\includegraphics[scale=0.06]{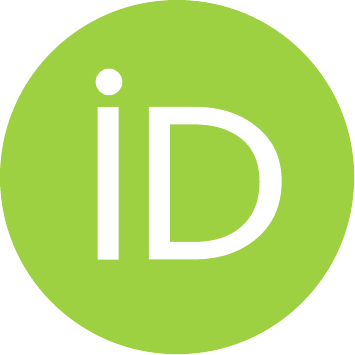}\hspace{1mm}David Baumgartner}\\
	\texttt{david.baumgartner@tugraz.at} \\
	\And
	\href{https://orcid.org/0000-0001-7576-0790}{\includegraphics[scale=0.06]{orcid.pdf}\hspace{1mm}G\"unter Brenn} \\
	\texttt{guenter.brenn@tugraz.at} \\
	\And
		\href{https://orcid.org/0000-0002-3974-5742}{\includegraphics[scale=0.06]{orcid.pdf}\hspace{1mm}Carole Planchette}\thanks{Corresponding author}  \\
	\texttt{carole.planchette@tugraz.at} \\
	\\
	Institute of Fluid Mechanics and Heat Transfer, Graz University of Technology, A-8010 Graz, Austria
	
}

\renewcommand{\shorttitle}{Universality of stretching separation}

\hypersetup{
pdftitle={Universality of stretching separation},
pdfsubject={Fluid mechanics, Milli and micro fluid mechanics},
pdfauthor={David Baumgartner, G\"unter Brenn, and Carole Planchette},
pdfkeywords={drop-drop collisions, drop-jet collisions, in-air microfluidics},
}

\begin{document}
\maketitle

\begin{abstract}
We develop a parameter-free model for the fragmentation of drops colliding off-center.  The prediction is excellent  over  a wide range of liquid properties. The so-called stretching separation is attributed to the extension of the merged drop above a critical aspect ratio of {3.25}. The evolution of this aspect ratio  {is influenced by the liquid viscosity} and can be interpreted via an energy balance. This approach is then adapted to drop-jet collisions, which we model as consecutive drop-drop collisions. {The fragmentation criterion is similar the one observed at drop-drop collisions}, while the evolution of the stretched jet aspect ratio is modified to account for the different flow fields and geometry. 

\end{abstract}


Collisions of two or more droplets of one single liquid in a gaseous environment is a  common phenomenon whose importance {motivated} many theoretical \citep{ref:Roisman2009a, ref:Roisman2009b}, numerical \citep{ref:Sun2015, ref:Moqaddam2016, ref:Huang2019, ref:Li2016} and experimental \citep{ref:Brenn2001, ref:Brenn2006, ref:Pan2009} studies. It may cause droplets to {permanently} coalesce and trigger rainfall \citep{ref:Jayaratne1964}. {It may also lead to  fragmentation and therefore to the formation of drops with different sizes and trajectories. These phenomena also occur in industrial processes, which produce or employ drops, such as  coating, injection, cooling,...\citep{ref:Brenn1996}. }
 The consequences depend on the application, {but can be severe. It may alter the active delivery from spray-dried particles or modify the intake of pesticide sprayed onto crops.  Thus, if not controlled, the collision outcomes must be predicted. }Results across multiple studies reveal that coalescence, bouncing, reflexive and stretching separations are the four main outcomes of drop-drop (D-D) collisions. {Which outcome is obtained depends on the collision parameters and on the liquid drop behaviour. The latter is characterised by the drop Ohnesorge number $Oh_d=\mu_d/\sqrt{\rho_d \sigma_d D_d}$,  which compares the relative importance of the capillary, spring-like contribution to the viscous, dashpot-like contribution. Here, $\mu_d$, $\rho_d$, $\sigma_d$ and  $D_d$ are the drop liquid viscosity, density, surface tension and the drop diameter, respectively. Note that $Oh_d$ is independent from $U$, the relative drop velocity.  Classically, regime maps are  represented for a fixed $Oh_d$ using the dimensionless impact parameter, which quantifies the collision eccentricity, and the drop Weber number $We_d=\rho_d D_d U^2/\sigma_d$, which represents the ratio of inertia over capillarity \citep{ref:Ashgriz1990, ref:Jiang1992, ref:Qian1997, ref:Saroka2015}. This letter focuses on off-centre drop collisions{, more precisely on the stretching separation causing the fragmentation of the otherwise permanently merged drop}. To date, and despite parameter adjustments, the existing models have limited validity ranges \citep{ref:Pan2019, ref:Rabe2010,  ref:Gotaas2007, ref:Finotello2017}}. {More recently, \citet{al-dirawi_2021} proposed a new model. Yet, its derivation is empirical and its applicability remains limited to $0.02<Oh_d<0.14$.} We {solve this problem} and {based on energy balance} establish a {unique}, general and robust model predicting the stretching separation for D-D collisions {valid at least for  $0.008<Oh_d<0.325$}.

{The concept of stretching separation is then extended to drops colliding with a continuous jet. These collisions, also called in-air microfluidics, create well-defined liquid structures \citep{ref:Baumgartner2019_ILASS}. Combined with a subsequent solidification step,  advanced  fibres  or  capsules are continuously produced with high precision and throughput \citep{ref:Visser2018}, opening the paths to new applications.} Yet, studies on drop-jet (D-J) collisions remain rather rare. \cite{ref:Chen2006} investigated the {out-of-plane} collisions of water drops with a water jet, followed by \cite{ref:Planchette2018}, who {worked  with immiscible liquid pairs on in-plane collisions}. In this case, the outcomes were classified according to the fragmentation of either the drops, the jet, both phases, or none. {While the drop fragmentation limit has been recently studied \citep{ref:Baumgartner2020, ref:Baumgartner2020_PRF}, the one of the jet remains poorly explained. We show that it is caused by its excessive  stretching and  can be  predicted by considering the D-J collisions  as a succession of off-center D-D collisions.}

\begin{figure}
\includegraphics[width=11cm]{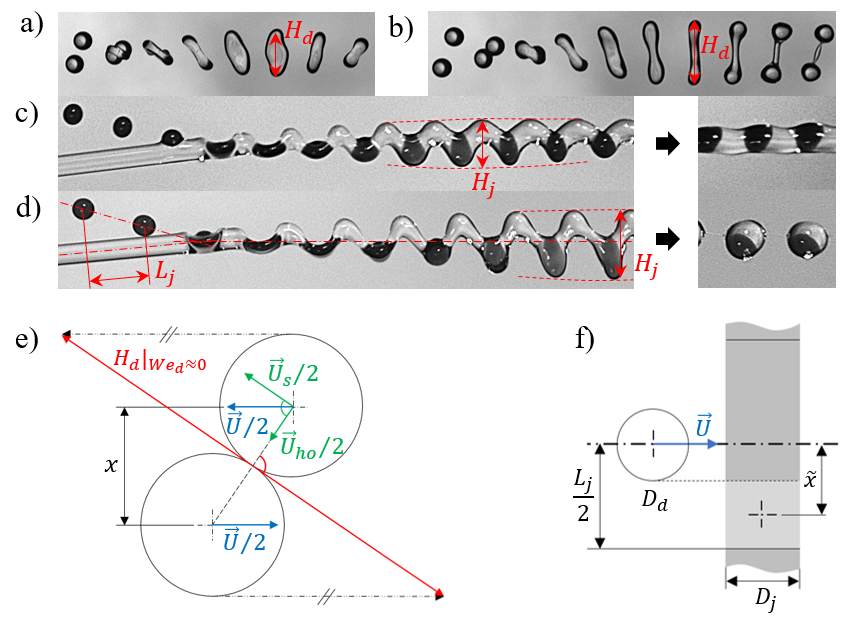}
\centering
\caption{\label{fig01} D-D collision in (a) coalescence ({$D_d=341 \mu m$}, X=0.41 $We_d=31.5$ $Oh_d=0.033$) and (b) stretching separation ({$D_d=340 \mu m$}, X=0.61 $We_d=31.3$ $Oh_d=0.033$). D-J collision in (c)  \textit{drops-in-jet}  ({$D_d=275 \mu m$}, $\tilde{X}=1.64$ $We_{d}=30$ $Oh_j=0.246$ and (d) \textit{capsules} ({$D_d=292 \mu m$}, $\tilde{X}=1.86$ $We_{d}=48$ $Oh_j=0.246$). Collision eccentricity for (e) D-D collisions: $X=x/D_d$ and (f) D-J collisions: $\tilde{X}=2\tilde{x}/D_j$.}
\end{figure}

While  different in nature, D-D and D-J collisions show similarities evidenced by Fig. \ref{fig01}.  {All pictures are obtained with a stroboscopic illumination at the drop production frequency (recording "standing" pictures) and are thus  made of tens to hundreds superimposed collisions.} Details about the experiments are given in Appendix \ref{appA}. Here and thereafter, subscript $d$ corresponds to drop liquid properties and  parameters while $j$ refers to those of the jet.

\begin{figure}
\includegraphics[width=\textwidth]{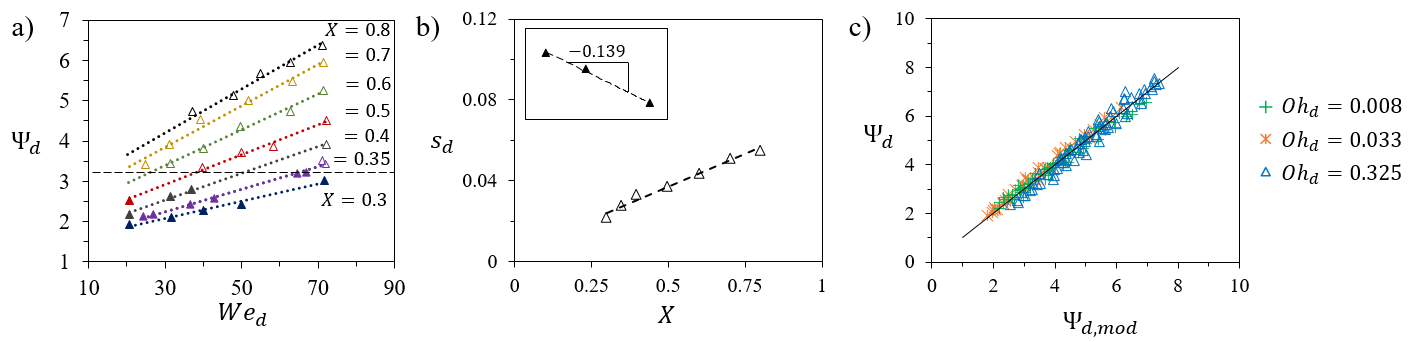}
\caption{\label{fig02} D-D collisions: (a) $\Psi_d$ as a function of $We_d$ { for $Oh_d=0.033$} and different $X$. Coalescence (full symbols) and separation (empty symbols). (b) $s_d= \partial \Psi_d/ \partial We_d$ {from (a)}. Inset: log($s_d/X$) against log($Oh_d$) {for all experiments ($0.008<Oh_d<0.325$)}. (c) Experiments versus model - Eq. (\ref{eq01_b}) for $0.008<Oh_d<0.325$.}
\end{figure}

Let us first focus on D-D collisions. Our model is based on two ingredients: {(i)} a fragmentation criterion, which corresponds to a critical value of $\Psi_d=max[H_d(t)/D_d]$, the maximum dimensionless drop extension (see Fig. \ref{fig01} (a) and (b)) and {(ii)} a function, which describes the variations of $\Psi_d$ with the liquid properties and collision parameters.
 We use three liquids:  water (W, 0.98mPa$\cdot$s, 72.5mN$\cdot$m$^{-1}$), an aqueous glycerol solution  (G, 5.1mPa$\cdot$s, 68.0mN$\cdot$m$^{-1}$) and silicone oil (SO, 18.5mPa$\cdot$s, 20.5mN$\cdot$m$^{-1}$) and vary the drop diameter between $175\mu m$ and $367\mu m$, covering $0.008 < Oh_d < 0.325$. The liquid properties are shown in Appendix \ref{appB}. The relative velocity $U$ ranges from $1.9 ms^{-1}$ to $5.4ms^{-1}$  corresponding to $15 < We_d < 265$.  $H_d(t)/D_d$  is then measured for several instants after the collision and as illustrated in Appendix \ref{appC}, fitted with a third-order polynomial to obtain $\Psi_d$, its maximum value. We verify that the separation threshold corresponds to a constant critical value of $\Psi_d$ of  {3.25} (dashed line in Fig \ref{fig02}(a)), in agreement with the numerical results of \cite{ref:Saroka2015} {and the recent experimental findings of \citet{al-dirawi_2021}}. {This provides the fragmentation criterion (i)}. {Note that, this value remarkably close to  $\pi$,  the theoretical one \citep{ref:Rayleigh1892}, is obtained by normalizing the critical extension with  the initial  diameter of one drop, as done by \citet{ref:Saroka2015} and \citet{ al-dirawi_2021}. Assuming a cylindrical shape and using volume conservation lead to a critical value of more than 5, well above the one of  Plateau-Rayleigh.}
 
 We then derive the evolution of $\Psi_d$ with the liquid properties and collision parameters{, i.e the second ingredient of our approach}. {We consider a purely geometric contribution $\Psi_d|_{We_d\approx0}$  and an inertial one $\Psi_d|_{We_d \neq 0}$}. The relative velocity $U$ is projected parallel ($\vec{U}_{ho}$) and normally ($\vec{U}_s$) to the  line connecting the two drop centers at the instant of contact (see  Fig. \ref{fig01}(e)). {The merged drop is elongated without inertial stretching along $\vec{U}_s$. This  elongation contributes to $\Psi_d$ in the form of a purely geometric term} $\Psi_d|_{We_d\approx0}=H_d|_{We_d\approx0}/D_d$, the normalized length of the red segment in Fig. \ref{fig01}(e). As shown in Appendix \ref{appD}. $\Psi_d|_{We_d\approx0}=(X+1)/\sqrt{1-X^2}$ and can  be well approximated by  $\Psi_d|_{We_d\approx0} \approx 2.7X+0.5$ for $0.3<X<0.8$. {The  inertial contribution is obtained by considering that some of the initial kinetic energy ($ = \pi \rho_d {D_d}^3 U^2 /24$)  is converted into surface energy of the stretched drop ($ \approx \sigma_d \pi H_d {D_d}$) providing at first order   $\Psi_d|_{We_d \neq 0} \approx f(X, Oh_d) We_d /24$ with $f(X, Oh_d)$ a function that accounts for the "relevant inertia". Here the "relevant inertia" is the one that causes the merged drop to stretch. Obviously, it corresponds to the inertia of } the almost unaffected drop portions that  continue on their initial trajectories. {A first correction must account for their reduced mass (or volume).} Neglecting strong  distortion, {each portion volume is given by  $V=(3X^2-2X^3)V_d$ with $V_d$  the volume of one drop, see Appendix \ref{appD}.  After linearization, we obtain $V/V_d \approx (1.4X-0.2)$ and thus, a linear variation of  $f(X, Oh_d)$ with $X$. To go further, the viscous losses must be estimated.}

Establishing the quantitative expression of this second correction is rather complex and we decide here to use the numerical findings of \cite{ref:Finotello2017}. The computed normalized remaining energy is replotted as a function of $X$ and $Oh_d$, see Appendix \ref{appF}. It   provides $f(X, Oh_d) \approx 0.31 X  {Oh}^{-0.18} $, in agreement with the expected linear variation of $f$ with $X$. As a consequence, we obtain $\Psi_d|_{We_d \neq 0} \approx 0.31/24 {Oh}^{-0.18} X We_d$ and thus the theoretical model:
\begin{equation}
\Psi_{d, th} =  \alpha_{d, th} Oh_d^{m_{th}}  X We_d + \beta_{d, th} X + \gamma_{d, th} \tag{1}
\label{eq01}
\end{equation}
The constants $\beta_{d, th}=2.7$ and $\gamma_{d, th}=0.5$ come from $\Psi_d|_{We_d \approx0}$ and are expected to be correct.   $\alpha_{d, th}= 0.31/24\approx0.013$ and $m_{th}=-0.18$ originate from  $\Psi_d|_{We_d \neq0}$ and thus, from the estimation of the "relevant inertia". Due to the uncertainty of the computed findings, the  approximation caused by linearization and the crude estimation of the stretched drop surface energy, these two constants are probably not quantitatively well predicted.  We  test the scaling of  our theoretical model (Eq. (\ref{eq01})) by plotting the experimental data $\Psi_{d}$ as a function of $We_d$ for different $X$. While these curves are obtained for all experiments ($0.008<Oh_d<0.325$), only those corresponding to $Oh_d=0.033$ are shown in  Fig. \ref{fig02}(a). Note that the line corresponding to {$\Psi_d=3.25$} separates well the coalescence (empty symbols) from the separation (full symbols). 
In agreement with Eq. (\ref{eq01}),  $\Psi_d$  increases linearly with $We_d$.  \cite{ref:Saroka2015}, who numerically studied water drop collisions, reported similar variations. 
Further, for a given {$Oh_d$ and fixed $X$}, the curve slopes  $s_d$ are linear in $X$ (Fig. \ref{fig02}(b)) as expected by Eq. (\ref{eq01}) (first term). Repeating the experiments with three liquids and thus three $Oh_d$, we find that $a_d=s_d/X$ is equal to 0.078 for W ($Oh_d=0.008$), 0.066 for G ($Oh_d=0.033$) and 0.047 for SO ($Oh_d=0.325$), and thus decreases with increasing  $Oh_d$. The  scaling $a_d \propto {Oh_d}^{-0.14}$ (see dashed line in the insert of Fig. \ref{fig02}(b)) is  reasonably well captured by Eq. (\ref{eq01}), which predicts $a_d \propto {Oh_d}^{-0.18}$.

To obtain a quantitative model, the experimental results $\Psi_{d}$ are fitted by $\Psi_{d, mod} =  \alpha_{d, mod} Oh_d^{m_{mod}}  X We_d + \beta_{d, mod} X + \gamma_{d, mod}$. There, only $m_{mod}$ and  $\alpha_{d, mod}$ are adjusted  while  $\beta_{d, mod}$  and $\gamma_{d, mod}$ are taken  equal to $\beta_{d, th}$  and $\gamma_{d, th}$, see Fig. \ref{fig02}(c). We obtain an excellent agreement with:
\begin{equation}
\Psi_{d, mod} =  0.041 Oh_d^{-0.128}  X We_d + 2.7 X + 0.5
\tag{2} \label{eq01_b}
\end{equation}

 The discrepancy between $m_{th}$ and $m_{mod}$ ($28\%$) could originate from the integration by \cite{ref:Finotello2017} of the losses over the whole process instead of the first instants, see Appendix \ref{appF} for details.  {The fit also provides $\alpha_{d, mod}=0.041$, while the theory gives $\alpha_{d, th}=0.013$. The difference (factor 3) can be explained by the crude estimation of the stretched drop surface.} All constants being the same for $0.008 < Oh_d < 0.325$, the validity of Eq. (\ref{eq01_b}) is very wide. {The agreement is also excellent while using the data of \citet{al-dirawi_2021} (not shown).  This indeed indicates that  the stretching separation is not purely inertial as previously reported \cite{al-dirawi_2021}. In fact, the authors probed less $We_d$ over a smaller range of $Oh_d$, which did not allow to identify the variations of $  \partial \Psi_d/ \partial X$ with $Oh_d$}. We then predict the separation threshold in the form of an ($X$, $We_d$) relation by fixing in Eq. (\ref{eq01_b}) $\Psi_{d, mod}$ to its critical value of {3.25}. 
 The results are compared to those of the literature, see Fig. \ref{fig04}(a). First of all,   for all considered $Oh_d$ the predicted thresholds (lines) perfectly match the experimental ones (symbols). We further observed a very good agreement to previously proposed models, which involved adjusted parameters while being limited to given $Oh_d$ \citep{ref:Jiang1992, ref:Ashgriz1990, ref:Finotello2017, ref:Gotaas2007}. We recall, that with our approach, no parameter is adjusted to cover collisions with $0.008<Oh_d<0.325$.

 Let us now apply these results to D-J collisions.  There, the drops are always made of G, the aqueous glycerol solution, while silicone oils of various viscosities are used for the jet.  The jet diameter is equal to $280\pm10\mu m$, resulting in jet Ohnesorge numbers $0.02 < Oh_j=\mu_j/\sqrt{\rho_j D_j \sigma_j} < 0.25$. The droplet diameter varies between $190 \mu m$ and $370 \mu m$, leading to diameter ratios $0.7 < \Delta=D_{d}/D_j < 1.3$. The relative impact velocity $\vec{U}$ (3 m$\cdot$s$^{-1} < |\vec{U}| < 8 m\cdot$s$^{-1}$)  is adjusted to be perpendicular to the jet axis. In former studies \citep{ref:Planchette2018, ref:Baumgartner2020, ref:Baumgartner2020_PRF}, the spatial period of the jet $L_j$ was  normalized by $D_j$  and used to build a pseudo-Rayleigh criterion. A critical value of 2 was found to roughly describe the jet fragmentation threshold in the limit of moderate jet viscosity.  Here, the analogy with D-D collisions requires the introduction of a new parameter to quantify the eccentricity of the successive collisions. As sketched in Fig. \ref{fig01}(f), these collisions involve a drop and the jet portions found before and after this drop. Thus, the distance between the center of mass of the drop and one jet portion is given by $\tilde{x}=(L_j+D_d)/4$. This distance is counted twice since the drop interacts with the jet sections before and after. Using $D_j$ for normalization, the equivalent impact parameter reads $\tilde{X}=(L_j+D_d)/(2D_j)$.
 
 We record several D-J collisions and define, similarly to  $H_d(t)/D_d$, the  dimensionless extension of the jet $=H_j(t)/D_j$  measured perpendicularly to the final drop-jet trajectory (Fig. \ref{fig01}(c) and (d)). Its temporal evolution is fitted by a third-order polynomial providing its maximum value $\Psi_j=max(H_j(t)/D_j)$, for an example see Appendix \ref{appC}.
\begin{figure}
\includegraphics[width=\textwidth]{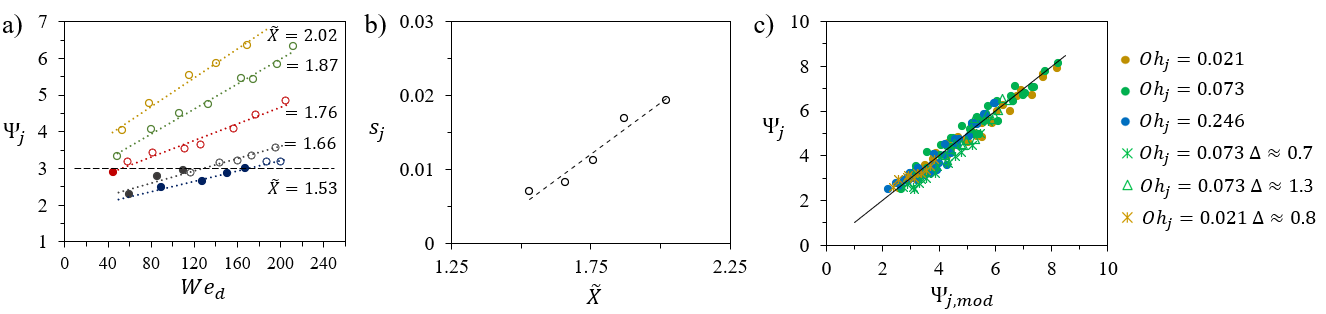}
\caption{\label{fig03} D-J collisions: (a) $\Psi_j$ against  $We_d$ for different $\tilde{X}$,  {$Oh_j=0.246$ and $\Delta=1.0$}. Coalescence (full symbols) and separation (empty symbols). (b) $s_j=\partial \Psi_j/ \partial We_d$  against $\tilde{X}$ {from (a)}. (c) Experiments versus model for - Eq. (\ref{eq02}) with $0.021<Oh_j<0.25$ and  $0.7<\Delta <1.3$.}
\end{figure}

We first confirm that the jet fragmentation  corresponds to a critical value of $\Psi_j$, see Fig. \ref{fig03}(a)). Interestingly, this critical value is {3.0, thus remarkably close to the one found for the D-D stretching separation} and underlines the relevance of our analogy.  We then plot for different $\tilde{X}$, the evolution of $\Psi_j$ with $We_{d}$ and evidence a linear dependency as for $\Psi_d$ for D-D collisions.  The curve slopes, $s_j$,  are again linear in $\tilde{X}$ and increase with {$Oh_j$}.  For D-J collisions,  the relevant {Ohnesorge number} is the one of the jet liquid, since {the viscous losses mainly take place in the interstitial jet portions, which are the most stretched}.
 We therefore propose to describe the jet stretching as: 
 \begin{equation}
\Psi_{j, mod} = \alpha_{j} Oh_j^n \tilde{X} We_{d} + \beta_j Oh_j^n \tilde{X} + \gamma_j
\tag{3} \label{eq02}
 \end{equation}
Here again, $\alpha_j$,  $\beta_j$ and $\gamma_j$ are constants. The first term accounts for  the drop inertia educed by viscous losses taking place in the jet only. The  last two terms correspond to geometrical effects.
As shown in Fig. \ref{fig03}(c), the agreement is again very good. The fit provides  -0.10 for the exponent $n$, close to -0.128 found for $m_{mod}$ and therefore supports the assumption that the viscosity (of drop and jet) plays despite different geometries, a similar role in both processes (D-D and D-J collisions, respectively). $\alpha_j$, $\beta_j$ and $\gamma_j$ are found to be +0.0066, +3.98, and -5.85, respectively. The slight deviation observed for D-J could have several origins. First of all, the system center of mass is approximated by the one of the jet, which slightly affects the collision inertia. Second, immiscible liquids are used, which modify the {flow field and thus the viscous losses. Due to the lack of existing data, Eq. (\ref{eq02}) could not yet be tested against results obtained with miscible liquids. It should definitively be done in future investigations.} Finally and despite its similarities, the process itself is  different. For D-J collisions, the system is continuous, and mainly shear occurs between the colliding elements. For D-D, the drop pairs constitute a close system, which can rotate around their center of mass, consuming part of the available inertia.

 To explain why, in contrast to $\Psi_d|_{We_{d}\approx0}$, $\Psi_j|_{We_{d}\rightarrow0}=\beta_j Oh_j^n \tilde{X} + \gamma_j$  is a function of $Oh_j$, it is useful to recall that Ohnesorge numbers can be seen as the ratio of a bulk motion  timescale, $t_{\mu}=\mu_jD_j/\sigma_j$ \citep{stone}, and an interfacial timescale, $t_{\sigma}=\sqrt{\rho_j {D_j}^3/\sigma_j}$ \citep{ref:Rayleigh1892}. At intermediate timescales, when $\Psi_j$ is measured, the morphology of the compound jet depends on their relative kinetics and therefore on the jet liquid properties via its Ohnesorge number. For high $Oh_j$, $t_{\mu}>t_{\sigma}$, the capillary effects are fast enough to significantly flatten the outer jet surface, leading to small $\Psi_j|_{We_{d}\rightarrow0}$. The contrary happens for small $Oh_j$. We also verify that   increasing  $L_j$ or $D_d$ as well as decreasing $D_j$ (thus increasing $\tilde{X}$) lead - as expected - to  
 greater $\Psi_j|_{We_{d}\rightarrow0}$. 
 
Note that, given the definition of $\tilde{X}$ and the value of $\beta_j$, $\gamma_j$ must be negative to represent separated successive and not overlapping drop collisions. We verify that  $\Psi_j|_{We_{j}\rightarrow0}>1$ in all experiments.

\begin{figure}
\includegraphics[width=11cm]{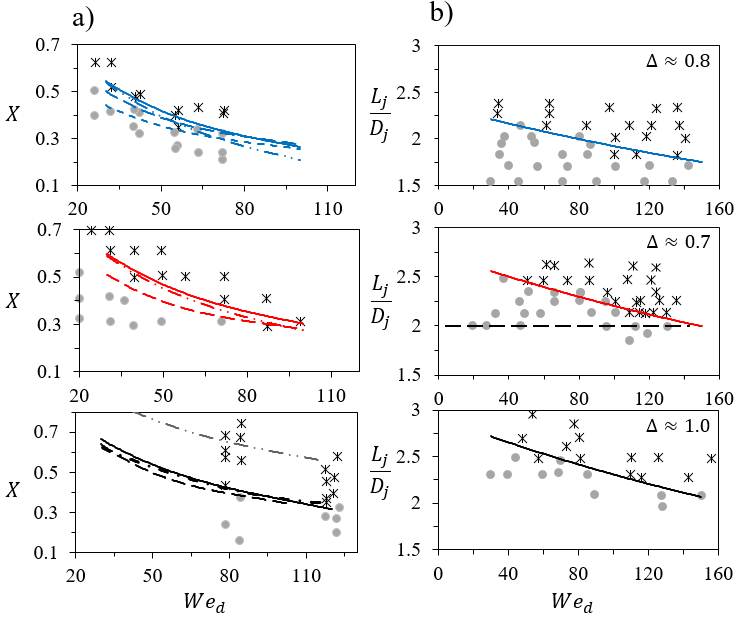}
\centering
\caption{\label{fig04} (a) D-D collisions: Separation transition for $Oh_d=0.008$ (top), $Oh_d=0.033$ (middle) and $Oh_d=0.325$ (bottom) with coalescence (circles) and separation (stars). Solid lines: {Eq. (\ref{eq01_b}) with {$\Psi_{d, mod}=3.25$}}, dashed lines: \cite{ref:Jiang1992}, dash-double-dotted lines: \cite{ref:Finotello2017}, dotted line: \cite{ref:Ashgriz1990} and dot-dashed line: \cite{ref:Gotaas2007}. (b) D-J collisions: Transition between continuous (circles) and fragmented jet (stars) for $Oh_j=0.021$ (top), $Oh_j=0.073$ (middle) and $Oh_j=0.246$ (bottom). Solid lines:  Eq. (\ref{eq02}) with { $\Psi_{j, mod}=3.0$. }Dashed line (middle): former criterion $L_j/D_j=2$. }
\end{figure}

As for D-D collisions, fixing $\Psi_{j, mod}$ to 3.0 in Eq. (\ref{eq02}) enables to predict the transition between continuous (circles) and fragmented jet (diamonds), see solid lines in Fig. \ref{fig04}(b). The agreement between the model (solid lines) and the experiments (symbols) is very good, significantly better than with the former criterion of $L_j/D_j \approx 2.0$ (horizontal dashed line). It is valid over a wide range of $Oh_j$ ($0.021 < Oh_j < 0.246$) and for different drop and jet diameters ($0.7<\Delta<1.3$) without adjusting any parameter. For additional results representing the separation transition at D-J collisions, and comparing the model and the experiments, see Appendix \ref{appG}.    

In conclusion, we have investigated off-centered  drop-drop and drop-jet collisions and found a universal  model for the transition  between coalescence and fragmentation caused by stretching separation. Our approach is based on (i) a simple transition criterion: a critical drop or jet extension of {3.25 or 3.0},  and (ii) the evolution of this drop or jet extension with the liquid properties and collision parameters. In contrast to other models of drop-drop collisions, our model remains valid for a wide range of Ohnesorge numbers - at least  over $0.008<Oh_d<0.325$ - without adjusting any parameter. For drop-jet collisions, our model is valid at least for $0.02<Oh_j<0.25$ and $0.7<\Delta<1.3$ with a precision going far beyond  the existing approach based on a critical $L_j/D_j$. The similarities between the collision morphologies, the transition criterion ({$\Psi_d=3.25$ and $\Psi_j=3.0$}), and the evolution of the maximum drop or jet extension (linear in $We_d$, linear in $X$ or $\tilde{X}$ and modulated by $Oh$ to the power of $m$ or $n$) underline the universality of our approach and of the so-called stretching separation. It could certainly be successfully applied to further situations, {such as miscible D-J collisions and beyond}.
\FloatBarrier

\section*{Acknowledgment}
We would like to thank the Austrian Science Fund (FWF) for the financial support under Grant No. P31064-N36.

\appendix
\section{Experimental set-up and problem parameters}\label{appA}

\begin{figure}
\centering
\includegraphics[width=8cm]{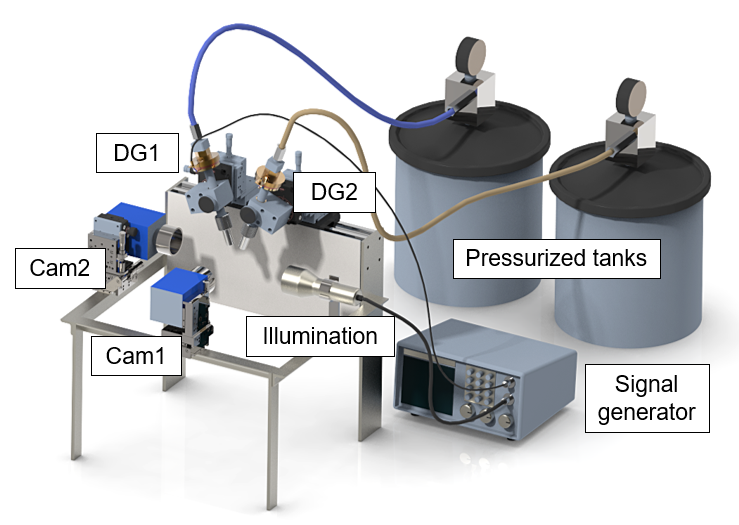}
\caption{\label{app_fig01} Experimental set-up for D-D and D-J collisions. }
\end{figure}

All drop-drop (D-D) and drop-jet (D-J) collisions are carried out with the same set-up depicted in  Fig. \ref{app_fig01}. For D-D collisions, both droplet generators \citep{ref:Brenn1996} are excited  with the same frequency $f_d$ (D-D; $5 kHz<f_d<22 kHz$ and D-J; $8 kHz<f_{d}<24 kHz$) while for D-J collisions, only one of the two generators (DG1) is activated, the second one being used as a simple orifice. An LED flash light (illumination) is synchronized with $f_d$  to generate frozen collision pictures. {Each picture therefore corresponds to tens to hundreds flashes of approximatively 100$ns$ each and is thus made of tens to hundreds superimposed collisions.} The liquids are supplied with two independent pressurized tanks. Two different cameras record collision pictures in front (Cam1) and orthogonal (Cam2) view. Pictures generated with Cam1 (camera resolution up to $4\mu m$ per pixel) are used for the measurements of $\Psi_d$ and $\Psi_j$.  Cam2 is primarily used to control the alignment in the collision plane of the drop and jet trajectories, which are adjusted with microtraverses.
\begin{figure}
\centering
\includegraphics[width=9cm]{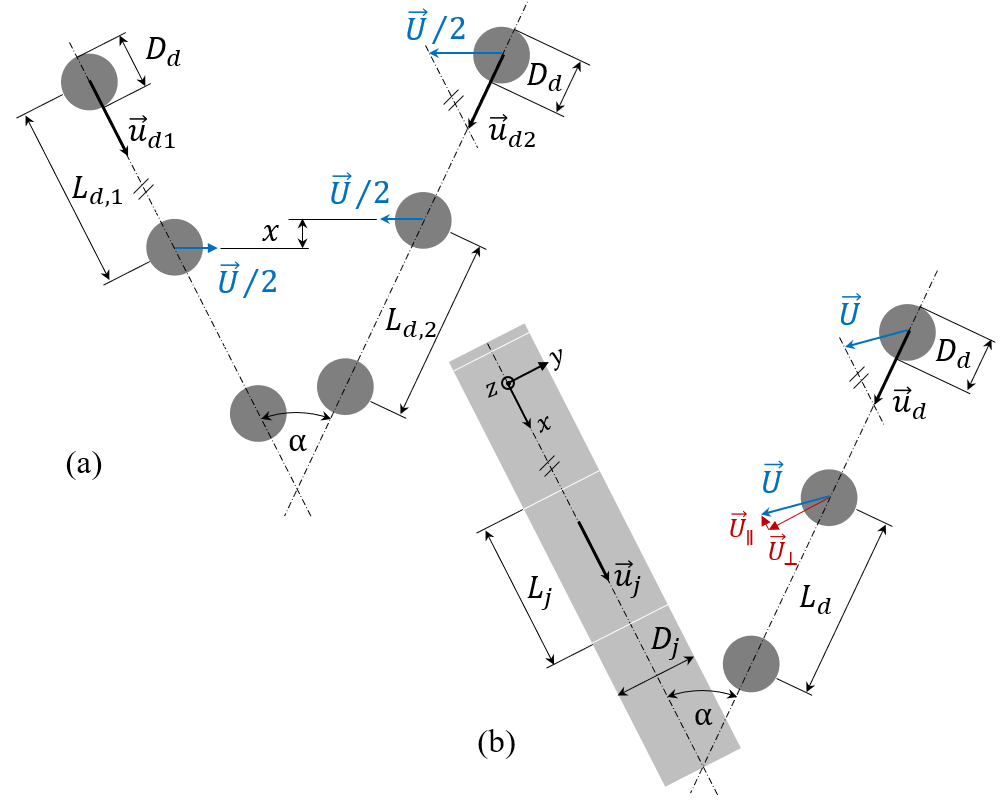}
\caption{\label{app_fig02} Geometrical and kinetic parameters of (a) D-D and (b) D-J collisions.}
\end{figure}

Fig. \ref{app_fig02} graphically shows the geometrical and kinetic parameters introduced for (a) D-D  and (b) D-J  collisions. All geometrical parameters are extracted from recorded pictures with the public-domain software \href{https://imagej.nih.gov/ij/}{ImageJ}. $D_d$ and $D_j$ stand for the drop and jet diameter, respectively. For D-D collisions, $L_{d,1}$ and $L_{d,2}$ correspond to the distance between two consecutive drops of the left and right droplet stream, and $\alpha$ ($18^\circ < \alpha < 50^\circ$) represents the collision angle. $X=x/D_d$ is the dimensionless impact parameter of the collision. It is defined as the distance between the drop centers normalized by their diameter.  The droplet velocities $\vec{u}_{d,1}$ and $\vec{u}_{d,2}$ are calculated with $\vec{L}_{d,1}f_d$ and $\vec{L}_{d,2}f_d$, respectively and vary between $3ms^{-1}$ and $10ms^{-1}$. The relative impact velocity $\vec{U}$ is then given by  $\vec{U}=\vec{u}_{d,2}-\vec{u}_{d,1}$.

For D-J collisions, $\alpha$ ($15^\circ < \alpha < 60^\circ$) is the collision angle and $L_{d}$  the distance between two consecutive drops. The spatial period of the jet and is defined as $L_j=|\vec{u}_j|/f_{d}$, where $|\vec{u}_j|$ is the absolute value of flow-rate equivalent jet velocity ($3.5 ms^{-1}<|\vec{u}_j|<11.5 ms^{-1}$). It is calculated from the measured mass flow rate knowing the liquid density and jet section. $\vec{u}_{d}=\vec{L}_{d}f_d$ is the drop velocity  and ranges between $5ms^{-1}$ and $13ms^{-1}$.  The relative impact velocity $\vec{U}=\vec{u}_{d}-\vec{u}_{j}$. Its parallel component $U_{||}=u_{d} cos(\alpha)-u_j$ is set close to zero, more precisely, in this work $U_{||}<0.1U$. Further information can also be found in \cite{ref:Baumgartner2020, ref:Baumgartner2020_PRF}.  

\section{Liquid properties}\label{appB}
 
The density $\rho$, the dynamic viscosity $\mu$ and the surface tension $\sigma$ (against the environment) of the liquids, as well as their abbreviation (Abbr.) and usage (Appl.) are shown in Table \ref{tab01}. Deionized water (W) was used as well as an aqueous glycerol solution (G) prepared with 50$\%$ mass fraction of glycerol ($\geq98$\%, Carl Roth GmbH, Germany). For the jet liquid,  silicone oils (SO-x) from Carl Roth GmbH (Germany) and IMCD South East Europe GmbH (Austria) were either used as received or mixed to obtain the desired viscosity (x in mPa$\cdot$s). Note that SO20 corresponds to the silicone oil used for D-D collisions. For D-J collisions, the drops were dyed with Indigotin 85 (E 132, BASF, Germany) at a concentration of 1 g/l. The dye was added to the aqueous glycerol solution before the liquid properties were measured, see Table \ref{tab01}.  The interfacial tension between jet and drop liquid was measured $\sigma_{int} = 32\pm 3 mNm^{-1}$, confirming the total wetting of the drops by the jet. 

\begin{table}
\caption{\label{tab01} The liquids abbreviation, usage and properties measured at an ambient temperature of $T=23\,\pm 2^\circ$C }
\centering
\begin{tabular}{ccccc}
\hline
\addlinespace[0.2cm]
 &  & Density & Dynamic viscosity & Surface tension \\
  Abbreviation & Usage & $\rho$ $(kg\,m^{-3})$ & $\mu$ $(mPa\,s)$& $\sigma$ $mN\,m^{-1}$ \\
  \addlinespace[0.2cm]
\hline 
\addlinespace[0.2cm]
W & Drop & $995\,\pm 2$ & $0.98\,\pm 0.02$ & $72.5\,\pm 0.5$  \\
G & Drop & $1125\,\pm 10$ & $5.10\,\pm 0.15$ & $68\,\pm 2$  \\
SO1 & Jet & $845\,\pm 2$ & $1.40\,\pm 0.05$ & $17\,\pm 0.5$  \\
SO3 & Jet & $887\,\pm 5$ & $2.73\,\pm 0.05$ & $18.5\,\pm 0.5$  \\
SO5 & Jet & $908\,\pm 5$ & $5.10\,\pm 0.10$ & $19.5\,\pm 0.5$  \\
SO20 & Drop/Jet & $949\,\pm 5$ & $18.50\,\pm 0.50$ & $20.5\,\pm 1$  \\
\addlinespace[0.2cm]
\hline
\end{tabular}
\end{table}

{\section{Measurement of $\Psi$ from $H(t)/D$}} \label{appC}
\FloatBarrier
\begin{figure}
\centering
\includegraphics[width=8cm]{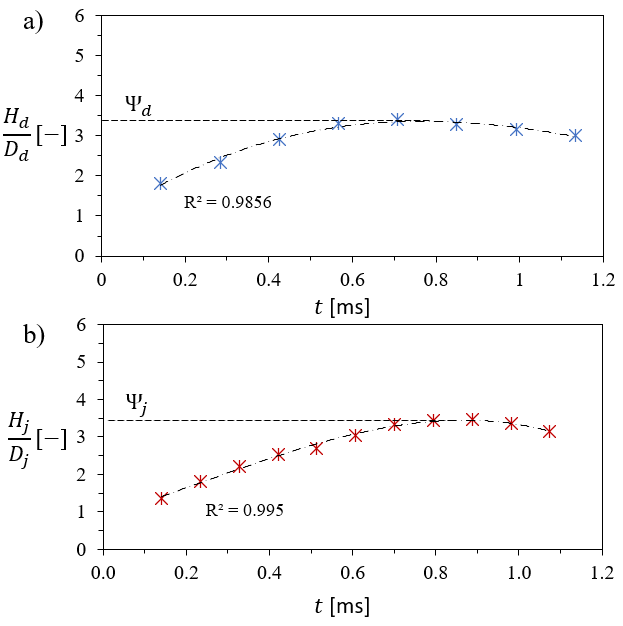}
\caption{\label{app_fig03} {(a) $H_d(t)/D_d$ for a D-D collision with $Oh_d=0.033$; $We_d=31.3$ and $X=0.61$ and (b) $H_j(t)/D_j$ for a D-J collision with  $Oh_j=0.246$; $We_d=81.3$ and $\tilde{X}=1.76$. }}
\end{figure}

{Fig. \ref{app_fig03} illustrates how $\Psi_d$ (respectively $\Psi_j$) are deduced from the temporal evolution of the drop (jet) distortion  $H_d(t)/D_d$ ($H_j(t)/D_j$). The latter is first experimentally measured (crosses) and then fitted  with a third order polynomial (dash dotted lines).   The regression coefficients  being always close to 1.0, the maximum of the fit function provides a very good estimation of  $\Psi_d$ and $\Psi_j$. }

\section{Non-interacting volume} \label{appD}

\begin{figure}
\centering
\includegraphics[width=9cm]{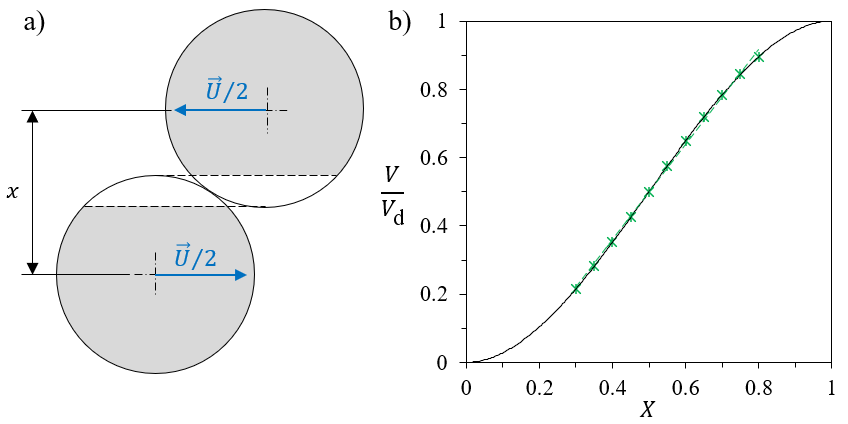}
\caption{\label{app_fig04} (a) Sketch showing (dark) the non-interacting drop volumes $V$ at the instant of contact. (b) $V/V_d$ as a function of X (black solid line). Green dashed line: the linear fit obtained over $0.3<X<0.8$.}
\end{figure}

For D-D collisions, assuming the drops neither significantly rotate nor deform, the non-interacting volume $V$ can be identified with the one of the external spherical cap (see dark areas in Fig. \ref{app_fig04}(a)). It yields $V=V_{d}(3X^2-2X^3)$, where $V_{d}=D_d^3\pi/6$ is the volume of one drop. The evolution of $V/V_{d}$ with $X$ is potted in Fig. \ref{app_fig04}b).  Within the range of our experiments ($0.3<X<0.8$), it can be estimated by the linear function $V/V_{d}=1.4X-0.2$,  green dashed line in Fig. \ref{app_fig04}(b).  The linear coefficient of determination $R^2=0.998$.

\section{Geometrical term} \label{appE}

\begin{figure}
\centering
\includegraphics[width=11cm]{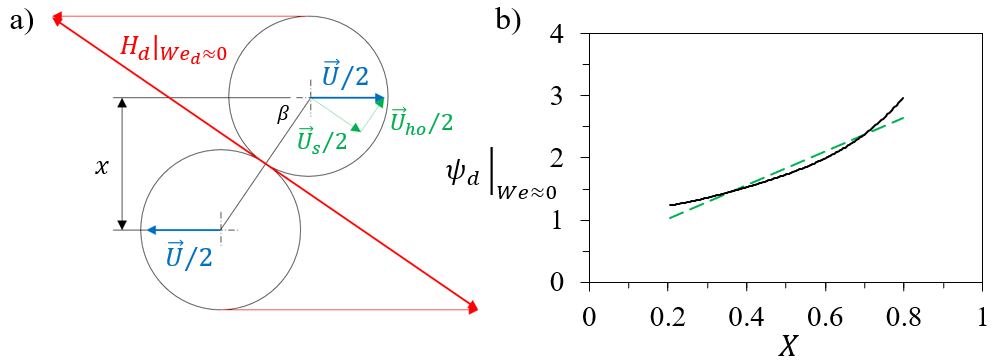}
\caption{\label{app_fig05} (a) Illustration of the instant of drop contact, showing $\vec{U}_{ho}$ and $\vec{U}_{s}$. (b) $\Psi_d|_{We_d\approx0}$ against X (black solid line) and the linear fit (green dashed line).}
\end{figure}

Here we derive the geometrical term of Eq. (\ref{eq01}) by considering a quasi-static collision. When the drops come into contact, their velocity can be projected along the line  joining their centers $\vec{U}_{ho}$ and perpendicularly to it $\vec{U}_{s}$, see Fig. \ref{app_fig05}(a). The former component $\vec{U}_{ho}$ is mostly responsible for "head-on" like deformation  - schematically leading to a lamella surrounded by a toroidal rim - and does not directly contribute to the stretching. In contrast, the latter one, $\vec{U}_{s}$, stretches the merged drop. The corresponding drop extension $H_d|_{We_d\approx0}$ can thus be estimated by the distance separating the drop extremities after projection along $\vec{U}_{s}$, see red line in Fig. \ref{app_fig05}(a). This length is given by $H_d|_{We_d\approx0}=(x+D_d)/\sqrt{1-X^2}$, providing, after normalization, $\Psi_d|_{We_d\approx0}= (X+1)/\sqrt{1-X^2}$, which is plotted in in Fig. \ref{app_fig05} (b) (black solid line).  Within the limits of the coalescence/stretching separation ($0.3<X<0.8$), it can be well approximated by a linear function (dashed green line). The latter corresponds to  $\Psi_d|_{We_d\approx0} = 2.7 X + 0.5$ with a regression coefficient of $R^2=0.946$. It perfectly corresponds to the geometrical term found in Eq. (\ref{eq01}).

{\section{Viscous losses}}\label{appF}

\begin{figure}
\centering
\includegraphics[width=9cm]{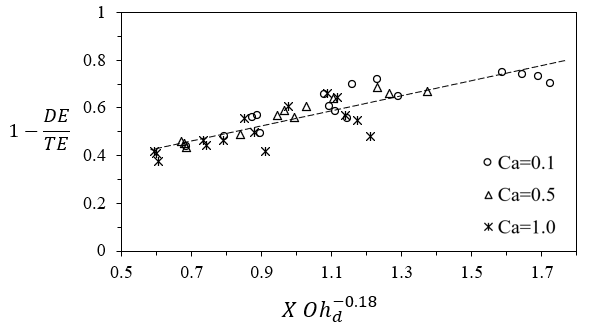}
\caption{\label{app_fig06} (1-DE/TE) - the energy, which is left for droplet stretching - as a function of X and $Oh_d$. The dashed line corresponds to $1-DE/TE=0.31X{Oh_d}^{-0.18}+0.25$.}
\end{figure}

To estimate the viscous losses taking place in the first part of D-D collisions, we make use of the numerical results of \cite{ref:Finotello2017}. More precisely, we replot the data of their Fig. 9. In its original form, it represents the variations of the dissipated energy $DE$ over the total initial energy $TE$ (approximated by the initial kinetic energy) as a function of the impact parameter $X$ for different Capillary numbers $Ca=\mu_dU/\sigma_d$ and Weber numbers - $X \in [0.3, 0.8]$ and $Oh_d=Ca_d/\sqrt{We_d} \in [0.01, 0.20]$. From these results, we derive the normalized remaining energy (1-$DE/TE$), which we plot  as a function of $X$ and a power of $Oh$ (see Fig. \ref{app_fig06}). The linear increase with $X$ is expected since the region of high dissipation rate is linearly growing for decreasing impact parameter, see \cite{ref:Finotello2017}.
We further choose to use  $Oh$ as it is commonly employed to weight the relative importance of viscosity {(dashpot behaviour)} and capillarity {(spring-like behaviour)}, the inertia being here indirectly accounted for via the normalization with the total initial energy which is approximated by the initial kinetic energy.  By doing so, the best fit reveals that the remaining energy, which is left for stretching the ligament between the drops scales as: $(1-DE/TE) \propto  X Oh_d^{-0.18}$. 
Note that there exists a constant loss so that for head-on collisions ($X \rightarrow 0$ ) approximately $25\%$ of the initial energy remains. This  is in agreement with the findings of \cite{ref:Planchette2017} who showed that for binary head-on collisions covering this range of $Oh$,  $35\%$ of the initial kinetic energy was left after the  drops deform into a lamella. The difference between $25\%$ and $35\%$ can be explained by the fact that Finotello et al. consider the duration of the entire collision process, while Planchette et al. (2017) only focus on its first phase. If most of the losses takes place during the first collision instants, dissipation continues to occur afterwards so that for low $Oh_d$ ($Oh_d \approx 0.02$, see Fig. 5a of \citep{ref:Finotello2017}) up to $23\%$ of the losses calculated by \cite{ref:Finotello2017} arise after $\varphi_d$ has reached its maximal extension $\Psi_d$. Note that for larger $Oh_d$ ($Oh_d \approx 0.1$, see Fig. 5b of \cite{ref:Finotello2017}), these subsequent losses are more limited, in the range of $6\%$. This may lead to an overestimation of the viscous losses in the case of small $Oh_d$ and could explain why in our model, which considers only the first phase of the collision until $\Psi_d$ is reached,  an exponent of $-0.128$ in Eq. (\ref{eq01}) provides a better agreement than the value of $-0.18$.

\FloatBarrier
\section{Additional results } \label{appG}
\begin{figure}
\centering
\includegraphics[width=11cm]{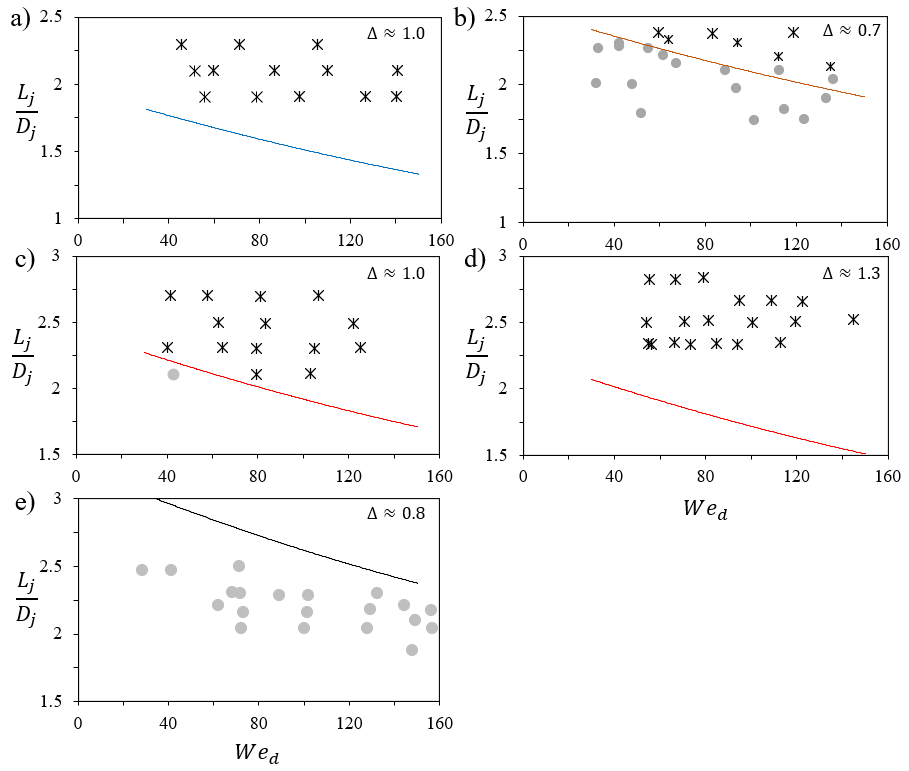}
\caption{\label{app_fig07} D-J collisions in coalescence (circles) and stretching separation (stars) for (a) $Oh_j=0.021$, (b) $Oh_j=0.040$, (c) and (d) $Oh_j=0.073$, and (e) $Oh_j=0.261$. Solid lines: Model Eq. (\ref{eq02}) with $\Psi_j=3.0$.}
\end{figure}
Fig. \ref{app_fig07} shows the results of D-J collision experiments for $Oh_j$ between 0.021 and 0.261, and different $\Delta=D_d/D_j$. The solid lines represent the separation transition according to the model of Eq. (\ref{eq02}) with the criterion $\Psi_j=3.0$.

\FloatBarrier

\bibliographystyle{unsrtnat}
\bibliography{references}

\end{document}